\documentclass[conference]{IEEEtran}
\IEEEoverridecommandlockouts

\usepackage[T1]{fontenc}
\usepackage{cite}
\usepackage{amsmath,amssymb,amsfonts,amsthm,bm,bbm,mathrsfs}
\usepackage{array,booktabs,multirow,makecell}
\usepackage{graphicx,xcolor}
\usepackage[ruled,vlined,linesnumbered]{algorithm2e}
\usepackage{setspace}
\usepackage{stfloats}
\usepackage{subfigure}
\usepackage{threeparttable}
\usepackage{dcolumn}
\usepackage[bookmarks=false]{hyperref}
\usepackage{ragged2e}

\newtheorem{rek}{Remark}

\graphicspath{{figures/}}
\begin{document}
	
	\title{Zero-Trust Bilateral Edge Service Trading with Deposit-Refund Regulation for Runtime Compliance}
	
	\author{
		\IEEEauthorblockN{
			Houyi Qi\IEEEauthorrefmark{1},
			Minghui Liwang\IEEEauthorrefmark{1},
			Zhipeng Cheng\IEEEauthorrefmark{2},
		Xiaoyu Xia\IEEEauthorrefmark{3}
		}
		\IEEEauthorblockA{
			\IEEEauthorrefmark{1}~Shanghai Research Institute for Intelligent Autonomous Systems, Tongji University, Shanghai, China\\
			\IEEEauthorrefmark{2}~School of Future Science and Engineering, Soochow University, Jiangsu, China\\
			\IEEEauthorrefmark{3}~ School of Computing Technologies, RMIT University, Melbourne, Australia\\
			Email: \{houyiqi@tongji.edu.cn, minghuiliwang@tongji.edu.cn, chengzp\_x@163.com, xiaoyu.xia@rmit.edu.au\}
		}
	}
	
	\maketitle
	
\begin{abstract}
	Privacy-sensitive edge services necessitate optimizing diverse-type resource scheduling to support trustworthy provisioning within a zero-trust security framework. However, existing studies rarely model how runtime compliance jointly affects bilateral clearing, ex-post settlement, and future seller eligibility in dynamic edge markets. To address this issue, we propose \textit{ZEBRIS}, a zero-trust bilateral edge service trading framework with deposit-refund regulation for privacy-sensitive services. Specifically, edge provisioning is modeled as a trading form of zero-trust-compliant service packages, where the buyer-side effective valuation captures service value, delay penalty, and privacy risk, while the seller-side effective ask incorporates resource and compliance costs. This yields a resource-aware positive-margin bilateral clearing mechanism under shared resource and security constraints. To discipline post-clearing moral hazard, we further design a capped deposit-refund settlement rule based on measurable runtime compliance and update each seller's future security posture according to realized compliance outcomes. ZEBRIS satisfies bilateral individual rationality and no-subsidy weak budget balance. Experiments demonstrate that ZEBRIS improves social welfare and compliance robustness while reducing service delay and privacy-risk-weighted cost over representative baselines.
\end{abstract}
	
	\begin{IEEEkeywords}
		Zero-trust, edge service trading, deposit-refund regulation, privacy-sensitive edge services
	\end{IEEEkeywords}
	
	\section{Introduction}
	With the evolution of next-generation communication systems, particularly 6G, together with the advancement of edge intelligence, an expanding class of privacy-sensitive and latency-critical applications, such as mobile intelligent assistants, immersive interaction, and real-time visual analytics, are increasingly dependent on proximate edge computing infrastructures to obtain low-latency communication-computation services~\cite{wang2024privacy,cheng2025privacy}. Users therefore continuously submit edge service requests to reduce local processing burden and improve service timeliness~\cite{cheng2025privacy,wang2025truthful}. However, practical multi-tenant edge environments are rarely fully trusted, extending service provisioning beyond conventional bandwidth, computation, and latency constraints. Runtime zero-trust enforcement, including continuous authentication, authorization, and monitoring~\cite{zheng2025eo,xu2025blockchain}, introduces verification overhead, compliance costs, and service uncertainty. Therefore, zero-trust security should be modeled not as a static trust label, but as a dynamic and measurable compliance factor that affects service feasibility, trading profitability, and post-execution accountability.
	
	Existing studies mainly fall into three separate directions, namely zero-trust security~\cite{zheng2025eo,xu2025blockchain}, auction-based edge service scheduling~\cite{wu2025effective,wang2025truthful}, and privacy-aware edge service optimization~\cite{cheng2025privacy,zhuang2025privacy}. Nevertheless, these lines of research are still largely developed in isolation, and they rarely provide a unified market mechanism in which runtime compliance simultaneously affects ex-ante bilateral clearing, ex-post settlement, and future seller eligibility. In particular, zero-trust studies mainly focus on spectrum trading or secure data sharing~\cite{zheng2025eo}, auction-based studies usually treat security as a static assumption~\cite{wang2025truthful}, and privacy-aware optimization does not explicitly model runtime compliance settlement or future seller-state feedback. As a result, privacy-sensitive edge service trading cannot be adequately modeled as a conventional one-shot resource assignment problem. Even after a transaction is cleared, service providers may reduce security efforts or deviate from promised quality levels, exposing buyers to privacy leakage and service degradation. Although deposit-refund mechanisms have been explored in zero-trust spectrum trading~\cite{zheng2025eo}, directly applying them to privacy-sensitive edge service markets is insufficient, because edge service trading further couples communication--computation assignment, privacy exposure, runtime compliance, and cross-round seller competitiveness.
	
	To address this, we propose \textit{ZEBRIS}, a \underline{ze}ro-trust \underline{b}ilateral trading framework with deposit-refund \underline{r}egulation for pr\underline{i}vacy-sensitive edge \underline{s}ervices. ZEBRIS trades \emph{zero-trust-compliant service packages} rather than bare resources. The buyer-side effective valuation captures service value, delay penalty, and privacy risk, while the seller-side effective ask incorporates resource and compliance costs. Based on positive-margin clearing, capped deposit-refund settlement, and seller-posture feedback, ZEBRIS forms a closed incentive-regulation loop between current runtime behavior and future market competitiveness. Main contributions are summarized as follows. 
	
	\noindent $\bullet$ To enable trustworthy provisioning in privacy-sensitive edge service markets, we formulate zero-trust bilateral service provisioning as a dynamic package-based trading problem, where communication--computation resources, delay requirements, privacy exposure, and runtime security compliance are jointly embedded into package feasibility and profitability. We then design \emph{ZEBRIS}, an online trading mechanism that forms a closed loop among ex-ante package clearing, ex-post capped deposit-refund settlement, and cross-round seller-posture evolution, thereby transforming runtime compliance from a static trust assumption into an economically regulated market factor.
	
	\noindent $\bullet$ We establish the key economic properties of \emph{ZEBRIS}, including bilateral individual rationality and no-subsidy weak budget balance under the well-designed midpoint pricing and deposit-capping rules. Experiments further demonstrate that \emph{ZEBRIS} improves social welfare, compliance robustness, service delay, and privacy-risk-weighted service cost over representative baselines.

\section{System Model and Problem Formulation}
\begin{figure}[!t] 
	\vspace{-0.0cm}
	\setlength{\abovecaptionskip}{-1 mm}
	\centering
	\includegraphics[width=1\columnwidth]{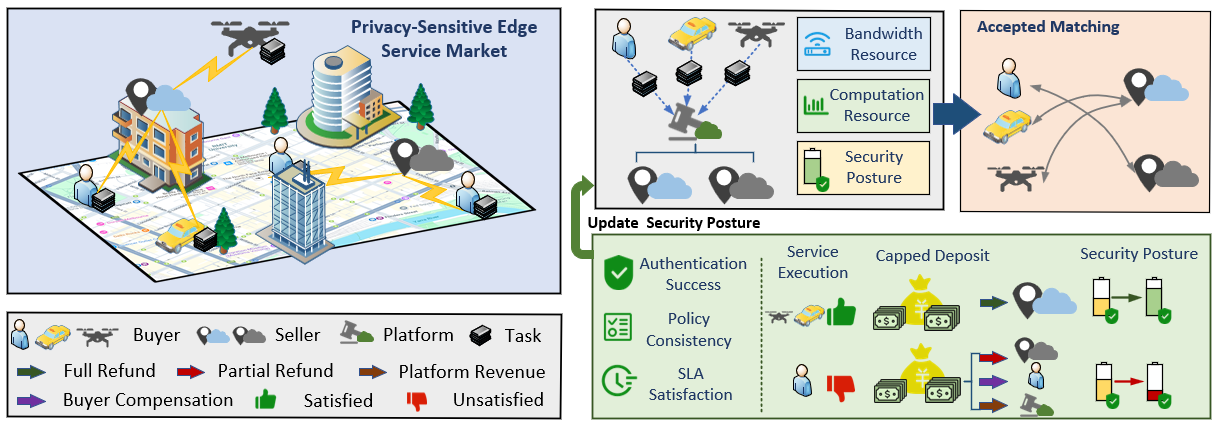}
	\caption{Framework and procedure of the proposed ZEBRIS.}
	\vspace{-0.6cm}\label{fig:system_model}
\end{figure}
As shown in Fig.~\ref{fig:system_model}, we consider a privacy-sensitive edge service market where buyers submit service tasks to edge sellers under platform coordination. Each seller offers zero-trust-compliant service packages characterized by bandwidth resource, computation resource, and security posture. For each buyer--seller pair, ZEBRIS evaluates candidate packages by jointly considering delay feasibility, privacy exposure, resource cost, and zero-trust compliance cost, and then performs positive-margin bilateral clearing to obtain the accepted matching.
After service execution, ZEBRIS measures runtime compliance through authentication success, policy consistency, and service level agreement (SLA) satisfaction. The capped deposit is then settled according to the measured compliance outcome, leading to full refund, buyer compensation, and platform revenue. The realized refund ratio is further fed back to update the seller's future security posture. Therefore, ZEBRIS establishes a closed-loop mechanism that integrates ex-ante package clearing, ex-post deposit-refund settlement, and cross-round seller-posture evolution for trustworthy privacy-sensitive edge service trading.
\subsection{Dynamic Zero-Trust Edge Service Market}

We consider a dynamic privacy-sensitive edge service market composed of a buyer set $\bm{U}$, whose members request edge services, a seller set $\mathcal{E}$, whose members provide bandwidth-computation resources and zero-trust-compliant packages, and a platform coordinator, while letting $\mathcal{T}=\{1,\ldots,T\}$ denote the trading horizon. In each round $t\in\mathcal{T}$, only a subset of buyers is active, denoted by $\mathcal{U}(t)\subseteq\bm{U}$.

On the demand side, each active buyer $u_i\in\mathcal{U}(t)$ submits a privacy-sensitive edge service request
$\mathcal{J}_i(t)=\bigl(L_i(t),\,C_i(t),\,D_i^{\max}(t),\,\ell_i(t),\,s_i^{\min}(t),\,v_i(t)\bigr),$
where $L_i(t)$ and $C_i(t)$ denote the input data size and required computation workload, respectively, $D_i^{\max}(t)$ is the delay deadline, $\ell_i(t)$ is the privacy sensitivity level, $s_i^{\min}(t)$ is the minimum required security level, and $v_i(t)$ is the buyer's gross valuation. Hence, each request jointly specifies communication/computation demand, timeliness requirement, and security/privacy requirement.
On the supply side, each seller $e_j\in\mathcal{E}$ is characterized by
$\mathcal{S}_j(t)=\bigl(B_j(t),\,F_j(t),\,q_j(t),\,\mathcal{Z}_j,\,a_j(t)\bigr),$
where $B_j(t)$ and $F_j(t)$ denote the available bandwidth and computation capacity, respectively, $q_j(t)\in[0,1]$ is the current security posture, $\mathcal{Z}_j\subseteq[0,1]$ is the feasible normalized verification-level set, and $a_j(t)$ is the base ask.

For each buyer--seller pair $(u_i,e_j)$ at round $t$, the platform considers a candidate package set $\mathcal{P}_{i,j}^{\mathrm{cand}}(t)$, where each candidate package is written as
$\mathcal{P}_{i,j}(t)=\bigl(b_{i,j}(t),\,f_{i,j}(t),\,z_{i,j}(t)\bigr)\in\mathcal{P}_{i,j}^{\mathrm{cand}}(t).$
Here, $b_{i,j}(t)$ and $f_{i,j}(t)$ denote the assigned bandwidth and computation resource, respectively, and $z_{i,j}(t)\in\mathcal{Z}_j$ denotes the selected normalized verification intensity. Let $x_{i,j}(t)\in\{0,1\}$ be the bilateral trading indicator, where $x_{i,j}(t)=1$ means that buyer $u_i$ is matched with seller $e_j$ at round $t$. In this paper, bilateral clearing refers to the platform's decision process that determines which buyer--seller pairs are accepted and which service packages are assigned, subject to positive-margin and resource-security feasibility constraints.

Unlike conventional edge auctions that trade resources, ZEBRIS trades zero-trust-compliant packages whose value depends on resource assignment, privacy exposure, and runtime compliance. Here, zero trust is modeled as continuous verification rather than a one-time trust label: each package must pass ex-ante security admission, undergo runtime monitoring, and be regulated through ex-post deposit-refund settlement. The seller posture $q_j(t)$ captures the platform-estimated compliance state and affects feasibility, privacy risk, delay overhead, compliance cost, and future competitiveness.

\subsection{Package Feasibility, Effective Valuation, and Effective Ask}

Based on the above market design, we next characterize how a candidate package affects service delay, privacy risk, and seller-side compliance cost. These factors determine both package feasibility and economic profitability under zero-trust enforcement.
For a candidate package $\mathcal{P}_{i,j}(t)$, let $\mathrm{SINR}_{i,j}(t)$ denote the signal-to-interference-plus-noise ratio (SINR) of the wireless link between buyer $u_i$ and seller $e_j$. The transmission rate is
$r_{i,j}(t)=b_{i,j}(t)\log_2\!\bigl(1+\mathrm{SINR}_{i,j}(t)\bigr).$
Then, the total service delay is modeled as
\begin{equation}
	D_{i,j}(t)=\frac{L_i(t)}{r_{i,j}(t)}+\frac{C_i(t)}{f_{i,j}(t)}+\vartheta_1 z_{i,j}(t)+\vartheta_2\bigl(1-q_j(t)\bigr),
	\label{eq:delay}
\end{equation}
where $\vartheta_1>0$ and $\vartheta_2>0$ are the delay coefficients associated with verification overhead and security-posture deficiency, respectively. To jointly capture the effects of verification intensity and seller posture, we define the package-level security compliance score as
$g_{i,j}(t)=g\bigl(z_{i,j}(t),q_j(t)\bigr)$,
where $g(\cdot)$ is nondecreasing in both arguments. For example, one simple instantiation is
$g(z,q)=\varpi z+(1-\varpi)q$ with $\varpi\in[0,1]$.
Accordingly, a candidate package is feasible only if
$D_{i,j}(t)\le D_i^{\max}(t), ~ g_{i,j}(t)\ge s_i^{\min}(t).$
These conditions ensure that the selected package simultaneously satisfies the buyer's timeliness requirement and zero-trust security requirement.
Beyond service delay, privacy exposure must also be explicitly quantified. The privacy risk experienced by buyer $u_i$ under seller $e_j$ is modeled as
\begin{equation}
	\xi_{i,j}(t)=\ell_i(t)\,\phi\bigl(z_{i,j}(t),q_j(t)\bigr),
	\label{eq:privacy}
\end{equation}
where $\phi(\cdot)$ is a nonnegative privacy-risk function that is decreasing in both $z_{i,j}(t)$ and $q_j(t)$. Thus, stronger verification and better seller posture reduce privacy exposure.
Meanwhile, zero-trust enforcement incurs an explicit seller-side compliance cost
$C_{i,j}^{\mathrm{zt}}(t)=\psi_1 z_{i,j}(t)+\psi_2\bigl(1-q_j(t)\bigr),$
where $\psi_1>0$ and $\psi_2>0$ are the cost coefficients associated with verification effort and posture deficiency, respectively.

Based on the above characterizations, we define the buyer-side effective valuation as
\begin{equation}
	\hat v_{i,j}\bigl(\mathcal{P}_{i,j}(t)\bigr)=v_i(t)-\alpha_i D_{i,j}(t)-\beta_i \xi_{i,j}(t),
	\label{eq:effv}
\end{equation}
where $\alpha_i>0$ and $\beta_i>0$ are the delay-penalty and privacy-risk-penalty coefficients, respectively. The seller-side effective ask is defined as
\begin{equation}
	\hat a_{i,j}\bigl(\mathcal{P}_{i,j}(t)\bigr)=a_j(t)+\kappa_j^\mathrm{B} b_{i,j}(t)+\kappa_j^\mathrm{F} f_{i,j}(t)+C_{i,j}^{\mathrm{zt}}(t),
	\label{eq:effa}
\end{equation}
where $\kappa_j^\mathrm{B}>0$ and $\kappa_j^\mathrm{F}>0$ are the seller's unit bandwidth and computation costs, respectively.
Accordingly, the bilateral trading margin is
\begin{equation}
	\Omega_{i,j}\bigl(\mathcal{P}_{i,j}(t)\bigr)=\hat v_{i,j}\bigl(\mathcal{P}_{i,j}(t)\bigr)-\hat a_{i,j}\bigl(\mathcal{P}_{i,j}(t)\bigr).
	\label{eq:margin}
\end{equation}
A positive margin means that the package remains beneficial after accounting for delay loss, privacy risk, resource cost, and compliance cost.

Let $\mathcal{P}_{i,j}^{\mathrm{fea}}(t)\subseteq \mathcal{P}_{i,j}^{\mathrm{cand}}(t)$ denote the feasible package set for pair $(u_i,e_j)$, namely the subset of candidate packages satisfying the above delay and minimum-security constraints. For each pair with $\mathcal{P}_{i,j}^{\mathrm{fea}}(t)\neq\varnothing$, the platform selects the best feasible package via
\begin{equation}
	\mathcal{P}_{i,j}^{\star}(t)\in \arg\max_{\mathcal{P}_{i,j}(t)\in\mathcal{P}_{i,j}^{\mathrm{fea}}(t)} \Omega_{i,j}\bigl(\mathcal{P}_{i,j}(t)\bigr),
	\label{eq:bestpkg}
\end{equation}
where $\mathcal{P}_{i,j}^{\star}(t)=\bigl(b_{i,j}^{\star}(t),\,f_{i,j}^{\star}(t),\,z_{i,j}^{\star}(t)\bigr).$
For notational convenience, define $\hat v_{i,j}^{\star}(t)\triangleq \hat v_{i,j}\bigl(\mathcal{P}_{i,j}^{\star}(t)\bigr) $, $ 	\hat a_{i,j}^{\star}(t)\triangleq \hat a_{i,j}\bigl(\mathcal{P}_{i,j}^{\star}(t)\bigr)$, and $\Omega_{i,j}^{\star}(t)\triangleq \Omega_{i,j}\bigl(\mathcal{P}_{i,j}^{\star}(t)\bigr) $.
If $\mathcal{P}_{i,j}^{\mathrm{fea}}(t)=\varnothing$, then pair $(u_i,e_j)$ is excluded from subsequent clearing.

\subsection{Runtime Compliance Settlement}
Ex-ante security-aware clearing cannot guarantee runtime compliance, since a winning seller may reduce verification effort, violate access-control policies, or fail to meet the promised service quality. To mitigate such post-clearing hazard under zero trust, we introduce an ex-post deposit-refund settlement mechanism based on measurable runtime compliance.

Specifically, for each accepted pair $(u_i,e_j)$ at round $t$, we evaluate realized runtime compliance from three complementary aspects, namely authentication success, policy consistency, and SLA satisfaction. Let $N_{i,j}^{\mathrm{req}}(t)$ and $N_{i,j}^{\mathrm{succ}}(t)$ denote the numbers of requested and successful authentication events, respectively, and let $N_{i,j}^{\mathrm{chk}}(t)$ and $N_{i,j}^{\mathrm{vio}}(t)$ denote the numbers of policy checks and detected policy violations, respectively. Moreover, let $D_{i,j}^{\mathrm{real}}(t)$ denote the realized end-to-end service delay after package execution. Then the three compliance scores are defined as
$A_{i,j}(t)=\frac{N_{i,j}^{\mathrm{succ}}(t)}{\max\{1,N_{i,j}^{\mathrm{req}}(t)\}},$
$G_{i,j}(t)=\left[1-\frac{N_{i,j}^{\mathrm{vio}}(t)}{\max\{1,N_{i,j}^{\mathrm{chk}}(t)\}}\right]_0^1,$
and
$S_{i,j}(t)=\left[1-\frac{\bigl(D_{i,j}^{\mathrm{real}}(t)-D_i^{\max}(t)\bigr)^+}{D_i^{\max}(t)}\right]_0^1,$
where $[x]_0^1\triangleq \min\{1,\max\{0,x\}\}$. As such, the ex-post settlement is grounded on measurable runtime outcomes rather than unverifiable behavioral assumptions.
The refund ratio is given by
\begin{equation}
	\rho_{i,j}(t)=\eta_1 A_{i,j}(t)+\eta_2 G_{i,j}(t)+\eta_3 S_{i,j}(t),
	\label{eq:refund}
\end{equation}
where $\eta_1,\eta_2,\eta_3\ge 0$ and $\eta_1+\eta_2+\eta_3=1$. Hence, by construction, $\rho_{i,j}(t)\in[0,1]$. A higher refund ratio indicates stronger realized compliance and better post-clearing service fulfillment.
To avoid excessive punishment that destroys seller participation, the deposit is capped as
\begin{equation}
	\Delta_{i,j}(t)=\min\!\left\{\mu_1 z_{i,j}^{\star}(t)+\mu_2\bigl(1-q_j(t)\bigr),\ \lambda\,\Omega_{i,j}^{\star}(t)\right\},
	\label{eq:deposit}
\end{equation}
where $\mu_1>0$ and $\mu_2>0$ are deposit coefficients associated with verification level and posture deficiency, respectively, and $\lambda\in(0,\frac{1}{2})$ is the deposit-cap ratio. This design ensures that the deposit remains large enough to discipline seller behavior while preserving participation incentives.

Let the refunded and forfeited deposits be
$\Gamma_{i,j}(t)=\rho_{i,j}(t)\Delta_{i,j}(t), ~
\Lambda_{i,j}(t)=\bigl(1-\rho_{i,j}(t)\bigr)\Delta_{i,j}(t),$
respectively. A fraction $\chi\in[0,1]$ of the forfeited deposit is returned to the buyer as compensation, and the remaining fraction is retained by the platform:
\begin{equation}
	C_{i,j}^{\mathrm{cmp}}(t)=\chi\,\Lambda_{i,j}(t), ~
	C_{i,j}^{\mathrm{plt}}(t)=\bigl(1-\chi\bigr)\Lambda_{i,j}(t).
	\label{eq:split}
\end{equation}
As a result, the proposed settlement rule not only disciplines seller-side post-clearing behavior, but also provides explicit buyer protection under weak realized compliance.

\subsection{Cross-Round Seller-Posture Evolution}
To further link current compliance with future competitiveness, we model cross-round seller-posture evolution.

For seller $e_j$, define the average refund ratio at round $t$ as
\begin{equation}
	\bar{\rho}_j(t)=
	\begin{cases}
		\dfrac{\sum_{u_i\in\mathcal{U}(t)} x_{i,j}(t)\rho_{i,j}(t)}{\sum_{u_i\in\mathcal{U}(t)} x_{i,j}(t)}, & \text{if } \sum_{u_i\in\mathcal{U}(t)} x_{i,j}(t)>0,\\[2ex]
		q_j(t), & \text{otherwise,}
	\end{cases}
	\label{eq:rhoavg}
\end{equation}
where the second case means that if seller $e_j$ is not selected in round $t$, its current posture remains the reference value. Then the seller posture evolves as
\begin{equation}
	q_j(t+1)=(1-\omega)q_j(t)+\omega\,\bar{\rho}_j(t), ~ \omega\in(0,1].
	\label{eq:posture}
\end{equation}

Thus, good compliance improves future posture, whereas weak compliance degrades future admissibility and competitiveness, since $q_j(t)$ affects package feasibility, privacy exposure, delay overhead, and compliance cost.

\subsection{Problem Formulation}

Building on the above modeling components, the platform first screens feasible packages for each buyer--seller pair and selects the best feasible package through \eqref{eq:bestpkg}. Define the feasible pair set at round $t$ as
$\mathcal{M}(t)\triangleq \left\{(u_i,e_j)\,\middle|\,\mathcal{P}_{i,j}^{\mathrm{fea}}(t)\neq\varnothing\right\}$.
After this screening step, each feasible pair $(u_i,e_j)$ is represented by its best package $\mathcal{P}_{i,j}^{\star}(t)$, and the platform only needs to determine the admissibility of the candidate matching. Accordingly, we use the following long-term social welfare maximization problem as a benchmark formulation to characterize the coupled clearing and seller-posture evolution:
\begin{equation}
	\mathcal{P}:\max_{x_{i,j}(t)}
	\sum_{t\in\mathcal{T}} \sum_{(u_i,e_j)\in\mathcal{M}(t)}
	x_{i,j}(t)\,\Omega_{i,j}^{\star}(t)
	\label{eq:obj}
\end{equation}
\begin{align}
	\mathrm{s.t.}\quad
	&\sum_{e_j:(u_i,e_j)\in\mathcal{M}(t)} x_{i,j}(t)\le 1, \tag{12a}\label{eq:c1}\\
	&\sum_{u_i:(u_i,e_j)\in\mathcal{M}(t)} x_{i,j}(t)b_{i,j}^{\star}(t)\le B_j(t), \tag{12b}\label{eq:c2}\\
	&\sum_{u_i:(u_i,e_j)\in\mathcal{M}(t)} x_{i,j}(t)f_{i,j}^{\star}(t)\le F_j(t), \tag{12c}\label{eq:c3}\\
	&x_{i,j}(t)\in\{0,1\}, \quad \forall (u_i,e_j)\in\mathcal{M}(t),  \tag{12d}\label{eq:c4}\\
	q_j(t&+1)=(1-\omega)q_j(t)+\omega\,\bar{\rho}_j(t),   \forall t\in\mathcal{T}\setminus\{T\}. \tag{12e}\label{eq:c5}
\end{align}

\noindent Here, payment, deposit, refund, and compensation are transfer terms among agents and are thus excluded from the social-welfare objective. Constraint \eqref{eq:c1} ensures that each buyer is matched with at most one seller in each round. Constraints \eqref{eq:c2} and \eqref{eq:c3} impose seller-side bandwidth and computation feasibility, respectively. Constraint \eqref{eq:c4} specifies binary trading decisions, and \eqref{eq:c5} captures cross-round seller-posture evolution. Since future runtime compliance outcomes and posture transitions cannot be fully observed before service execution, directly solving $\mathcal{P}$ as an offline clairvoyant problem is impractical. Problem $\mathcal{P}$ clarifies the coupled decision structure, while \textit{ZEBRIS} implements it through round-wise clearing based on current requests, seller states, and feasible packages.
\section{Proposed ZEBRIS}
\subsection{Round-Wise Clearing and Utility Settlement}

We next develop the proposed \textit{ZEBRIS}, with its core idea to convert dynamic zero-trust-constrained service trading into a resource-aware positive-margin bilateral clearing problem. Instead of comparing raw bids and asks, the platform compares the pair-wise best effective valuation and effective ask under feasible packages. In this way, delay loss, privacy risk, and runtime compliance cost are internalized into the clearing criterion itself.
We define the positive-margin feasible pair set as
$\mathcal{L}(t)=\{(u_i,e_j)\in\mathcal{M}(t)\mid \Omega_{i,j}^{\star}(t)>0\}$.
Only pairs in $\mathcal{L}(t)$ are eligible for clearing. Let
$\bm{X}(t)=\{x_{i,j}(t)\}_{(u_i,e_j)\in\mathcal{L}(t)}$
be the clearing outcome at round $t$. The feasible outcome set is
\begin{equation}{\small
	\begin{aligned}
		\mathcal{X}^{\mathrm{feas}}(t)\hspace{-1mm}=\hspace{-1mm}
		\Bigl\{\bm{X}(t)\big| \eqref{eq:c1}-\eqref{eq:c3}
		\text{ hold over } \mathcal{L}(t) \text{ at round }t\Bigr\}.
	\end{aligned}}
	\label{eq:Xfeas}
\end{equation}
Accordingly, the round-wise bilateral clearing problem is
\begin{equation}
	\max_{\bm{X}(t)\in\mathcal{X}^{\mathrm{feas}}(t)}
	\sum_{(u_i,e_j)\in\mathcal{L}(t)}
	x_{i,j}(t)\,\Omega_{i,j}^{\star}(t).
	\label{eq:roundclear}
\end{equation}
This problem serves as the round-wise clearing objective. To avoid myopic greedy admission, \textit{ZEBRIS} adopts a resource-discretized DP-based clearing rule over the reduced candidate graph induced by pair-wise representative packages. Each positive-margin pair $(u_i,e_j)$ is associated with a value-resource tuple $\bigl(\Omega_{i,j}^{\star}(t), b_{i,j}^{\star}(t), f_{i,j}^{\star}(t)\bigr)$. The DP sequentially scans candidate pairs in $\mathcal{L}(t)$, and its state records the processed pair index, the already matched buyers, and the remaining discretized bandwidth-computation resources of sellers. For each candidate pair, the transition either rejects it or accepts it if buyer-side exclusiveness and seller-side residual resource constraints are satisfied. The accepted transition increases the objective by $\Omega_{i,j}^{\star}(t)$ and reduces the corresponding seller resources by $b_{i,j}^{\star}(t)$ and $f_{i,j}^{\star}(t)$. This reduced-space design does not search over all raw packages jointly, but it preserves resource-aware positive-margin clearing over the selected representative packages with controllable overhead.

Once a buyer--seller pair is accepted, the platform proceeds to economic settlement. For each accepted trade $(u_i,e_j)$, the platform adopts a midpoint pricing rule
\begin{equation}
	p_{i,j}(t)=\frac{\hat v_{i,j}^{\star}(t)+\hat a_{i,j}^{\star}(t)}{2}.
	\label{eq:price}
\end{equation}
Based on this price, the per-trade buyer and seller utilities are respectively given by\footnote{The buyer utility in \eqref{eq:trade_utility} is a monetary settlement utility. Weak runtime compliance is not treated as better service experience, but is partially compensated by $C_{i,j}^{\mathrm{cmp}}(t)$ and separately evaluated through compliance and service-quality metrics.}
\begin{equation}
	\begin{aligned}
		U_{i,j}^{\mathrm{B}}(t)
		&=\hat v_{i,j}^{\star}(t)-p_{i,j}(t)+C_{i,j}^{\mathrm{cmp}}(t),\\
		U_{i,j}^{\mathrm{S}}(t)
		&=p_{i,j}(t)-\hat a_{i,j}^{\star}(t)-\Lambda_{i,j}(t).
	\end{aligned}
	\label{eq:trade_utility}
\end{equation}
Accordingly, the aggregate utilities are obtained by summing the corresponding per-trade terms over accepted pairs. Specifically, the buyer utility is
$U_i^{\mathrm{B}}(t)=\sum_{e_j:(u_i,e_j)\in\mathcal{L}(t)}x_{i,j}(t)U_{i,j}^{\mathrm{B}}(t)$,
the seller utility is
$U_j^{\mathrm{S}}(t)=\sum_{u_i:(u_i,e_j)\in\mathcal{L}(t)}x_{i,j}(t)U_{i,j}^{\mathrm{S}}(t)$,
and the platform revenue is
$U^{\mathrm{P}}(t)=\sum_{(u_i,e_j)\in\mathcal{L}(t)}x_{i,j}(t)C_{i,j}^{\mathrm{plt}}(t)$.

\subsection{Algorithm Summary}

\begin{algorithm}[b!]
	{\footnotesize
		\setstretch{1}
		\caption{Proposed ZEBRIS}
		\label{alg:ztbdr}
		\KwIn{Trading horizon $\mathcal{T}$, buyer requests $\{\mathcal{J}_i(t)\}$, seller states $\{\mathcal{S}_j(t)\}$, candidate package sets $\{\mathcal{P}_{i,j}^{\mathrm{cand}}(t)\}$.}
		\KwOut{Accepted trades, assigned packages, payments, settlements, and updated seller postures.}
		
		\For{each round $t\in\mathcal{T}$}{
			Initialize seller-side bandwidth and computation capacities by $B_j(t)$ and $F_j(t)$\;
			
			\For{each buyer--seller pair $(u_i,e_j)$ with $u_i\in\mathcal{U}(t)$ and $e_j\in\mathcal{E}$}{
				Enumerate candidate packages in $\mathcal{P}_{i,j}^{\mathrm{cand}}(t)$\;
				Discard infeasible packages violating delay or minimum-security requirements\;
				\If{$\mathcal{P}_{i,j}^{\mathrm{fea}}(t)\neq\varnothing$}{
					Obtain the best feasible package $\mathcal{P}_{i,j}^{\star}(t)$ and its margin $\Omega_{i,j}^{\star}(t)$\;
				}
			}
			
			Construct the positive-margin candidate set
			$\mathcal{L}(t)=\left\{(u_i,e_j)\in\mathcal{M}(t)\ \middle|\ \Omega_{i,j}^{\star}(t)>0\right\}$\;
			Discretize seller-side bandwidth and computation capacities\;
			
			Apply DP-based clearing over the reduced candidate set $\mathcal{L}(t)$ to obtain accepted trades $\{x_{i,j}(t)\}$ under buyer-side exclusiveness and seller-side resource constraints\;
			
			\For{each accepted pair $(u_i,e_j)$}{
				Assign $\mathcal{P}_{i,j}^{\star}(t)$\;
				Compute payment $p_{i,j}(t)$ by \eqref{eq:price}\;
				Compute capped deposit $\Delta_{i,j}(t)$ by \eqref{eq:deposit}\;
			}
			
			\For{each accepted pair $(u_i,e_j)$}{
				Execute the service package\;
				Measure $A_{i,j}(t)$, $G_{i,j}(t)$, and $S_{i,j}(t)$\;
				Compute refund ratio $\rho_{i,j}(t)$ by \eqref{eq:refund}\;
				Compute forfeited deposit $\Lambda_{i,j}(t)$ and settlement terms by \eqref{eq:split}\;
			}
			
			\For{each seller $e_j\in\mathcal{E}$}{
				Compute $\bar{\rho}_j(t)$ by \eqref{eq:rhoavg}\;
				Update $q_j(t+1)$ by \eqref{eq:posture}\;
			}
		}
	}
\end{algorithm}

As summarized in Alg.~\ref{alg:ztbdr}, \textit{ZEBRIS} performs six coupled operations in each round: feasible package screening, pair-wise best-package identification, DP-based positive-margin clearing, midpoint pricing with capped deposit assignment, ex-post compliance settlement, and seller-posture update. Specifically, the platform first identifies the best feasible package for each buyer--seller pair and constructs the positive-margin candidate set. It then applies the resource-discretized DP-based clearing rule over the reduced candidate graph to determine accepted trades under buyer-side exclusiveness and seller-side bandwidth-computation constraints. After service execution, the platform measures realized compliance, settles deposit refunds, and updates seller postures for the next round.

The per-round overhead mainly comes from feasible-package screening and DP-based clearing. Let 
$P_{\max}=\max |\mathcal{P}_{i,j}^{\mathrm{cand}}(t)|$ and let $|\mathcal{S}(t)|$ denote the discretized resource-state size. Since each candidate package can be evaluated in constant time, package screening costs 
$\mathcal{O}(|\mathcal{U}(t)||\mathcal{E}|P_{\max})$, and DP-based clearing costs 
$\mathcal{O}(|\mathcal{U}(t)||\mathcal{E}||\mathcal{S}(t)|)$. Therefore, the per-round complexity is
$\mathcal{O}(|\mathcal{U}(t)||\mathcal{E}|P_{\max}+|\mathcal{U}(t)||\mathcal{E}||\mathcal{S}(t)|)$,
where the discretization granularity controls the tradeoff between clearing accuracy and online overhead.

\subsection{Key Properties}
To demonstrate the economic soundness of ZEBRIS, we establish the following key properties.

\vspace{1mm}\noindent\textbf{Proposition 1 (Pre-settlement bilateral individual rationality).}
For any accepted trade $(u_i,e_j)$ with $\Omega_{i,j}^{\star}(t)>0$, both the buyer and the seller obtain strictly positive pre-settlement utility under the pricing rule in \eqref{eq:price}.

\noindent\emph{Proof.}
For any accepted trade $(u_i,e_j)$, we have
$\Omega_{i,j}^{\star}(t)=\hat v_{i,j}^{\star}(t)-\hat a_{i,j}^{\star}(t)>0$.
By the midpoint pricing rule in \eqref{eq:price},
$U_{i,j}^{\mathrm{B,pre}}(t)=\hat v_{i,j}^{\star}(t)-p_{i,j}(t)=\frac{1}{2}\Omega_{i,j}^{\star}(t)>0$
and
$U_{i,j}^{\mathrm{S,pre}}(t)=p_{i,j}(t)-\hat a_{i,j}^{\star}(t)=\frac{1}{2}\Omega_{i,j}^{\star}(t)>0$.
Thus, both sides obtain strictly positive utilities before ex-post settlement. \hfill$\square$

\vspace{1mm}\noindent\textbf{Proposition 2 (Ex-post seller individual rationality under deposit capping).}
If the deposit is set according to \eqref{eq:deposit} with $\lambda\in(0,\frac{1}{2})$, then every accepted trade brings strictly positive final utility to the corresponding seller, i.e., $U_{i,j}^{\mathrm{S}}(t)>0$ for each accepted pair $(u_i,e_j)$.

\noindent\emph{Proof.}
Since $\rho_{i,j}(t)\in[0,1]$, we have $\Lambda_{i,j}(t)\le \Delta_{i,j}(t)$. Using \eqref{eq:trade_utility},
$U_{i,j}^{\mathrm{S}}(t)=\frac{1}{2}\Omega_{i,j}^{\star}(t)-\Lambda_{i,j}(t)\ge \frac{1}{2}\Omega_{i,j}^{\star}(t)-\Delta_{i,j}(t).$
By \eqref{eq:deposit}, $\Delta_{i,j}(t)\le \lambda\,\Omega_{i,j}^{\star}(t)$. Thus,
$U_{i,j}^{\mathrm{S}}(t)\ge \left(\frac{1}{2}-\lambda\right)\Omega_{i,j}^{\star}(t)>0,$
because $\lambda\in(0,\frac{1}{2})$ and the accepted trade has positive margin. Hence, every accepted trade yields strictly positive per-trade seller utility. Since a seller's aggregate utility is obtained by summing its per-trade utilities over all accepted trades, the aggregate utility is nonnegative and becomes positive whenever at least one trade is accepted. \hfill$\square$

\vspace{1mm}\noindent\textbf{Proposition 3 (Dynamic compliance discipline).}
For any seller $e_j$, the posture update in \eqref{eq:posture} rewards above-reference runtime compliance and penalizes below-reference runtime compliance. Specifically, if $\bar{\rho}_j(t)>q_j(t)$, then $q_j(t+1)>q_j(t)$; if $\bar{\rho}_j(t)<q_j(t)$, then $q_j(t+1)<q_j(t)$; and if $\bar{\rho}_j(t)=q_j(t)$, then $q_j(t+1)=q_j(t)$.

\noindent\emph{Proof.}
From \eqref{eq:posture}, we have
$q_j(t+1)-q_j(t)=\omega\bigl(\bar{\rho}_j(t)-q_j(t)\bigr)$.
Since $\omega\in(0,1]$, the sign of $q_j(t+1)-q_j(t)$ is the same as the sign of $\bar{\rho}_j(t)-q_j(t)$. Therefore, above-reference realized compliance improves future posture, whereas below-reference realized compliance decreases future posture. Since $q_j(t)$ further affects package feasibility, privacy exposure, delay overhead, and compliance cost, the update links current runtime compliance to future market competitiveness. \hfill$\square$

\vspace{-1.5mm}
\begin{rek}
	\textbf{Buyer protection and weak budget balance.}
	For every accepted trade $(u_i,e_j)$, the monetary settlement utility of the buyer satisfies
	$U_{i,j}^{\mathrm{B}}(t)=\frac{1}{2}\Omega_{i,j}^{\star}(t)+\chi\,\Lambda_{i,j}(t)\ge \frac{1}{2}\Omega_{i,j}^{\star}(t)>0.$
	The compensation term should be interpreted as monetary protection against weak runtime compliance, rather than as an improvement of the realized service experience. Moreover, the platform only redistributes the forfeited deposit and retains $(1-\chi)\Lambda_{i,j}(t)\ge 0$, so it never needs to inject external subsidy. Hence, \textit{ZEBRIS} is no-subsidy and weakly budget balanced.
\end{rek}

	\section{Evaluation}
	
	We conduct simulations to evaluate the effectiveness of \textit{ZEBRIS}. All experiments are implemented in Python 3.10 on a 12th Gen Intel Core i9-12900H processor.
	
\subsection{Simulation Setup, Baselines, and Metrics}

We consider a dynamic privacy-sensitive edge service market with $|\mathcal{E}|=6$ sellers, average aggregate bandwidth of $48$ MHz, average aggregate computation capacity of $150\times10^9$ cycles/s, and $180$ trading rounds per episode. For each active buyer $u_i\in\mathcal{U}(t)$, the request tuple $\mathcal{J}_i(t)=\bigl(L_i(t),C_i(t),D_i^{\max}(t),\ell_i(t),s_i^{\min}(t),v_i(t)\bigr)$ is generated with $L_i(t)\in[0.15,0.95]$ MB, $C_i(t)\in[0.10,1.00]\times10^9$ cycles~\cite{qi2026Bank}, $D_i^{\max}(t)\in[0.25,0.90]$ s, $\ell_i(t)\in[0.20,1.00]$, $s_i^{\min}(t)\in[0.40,0.90]$, and $v_i(t)\in[8,20]$. Buyer activation follows Bernoulli trials with probabilities calibrated\cite{qi2026Bank} from the Chicago taxi trips dataset~\cite{chicago_taxi_trips_2013}.
Other parameters are set as follows~\cite{qi2026Bank,3GPP2022,wu2025effective}: $B_j(t)\in[6,10]$ MHz, $F_j(t)\in[18,32]\times10^9$ cycles/s, $q_j(t)\in[0.50,0.92]$, $a_j(t)\in[2,6]$, $\alpha_i\in[3,6]$, $\beta_i\in[2,5]$, $\kappa_j^\mathrm{B}\in[0.08,0.18]$, $\kappa_j^\mathrm{F}\in[0.10,0.22]$, $(\eta_1,\eta_2,\eta_3)=(0.35,0.30,0.35)$, $\chi=0.70$, and $\lambda=0.40$. We set $\phi(z,q)=(1-z)(1-q)$, generate $\mathcal{P}_{i,j}^{\mathrm{cand}}(t)$ by discretizing bandwidth, computation, and verification intensity, and average all results over 50 independent Monte Carlo runs.
Runtime compliance outcomes are generated by a deposit-aware effort level $\epsilon_{i,j}(t)=\sigma\!\left(\tau_0+\tau_1 q_j(t)+\tau_2\frac{\Delta_{i,j}(t)}{\hat a_{i,j}^{\star}(t)}\right)$, where $\sigma(\cdot)$ is the sigmoid function\footnote{The effort level only parameterizes stochastic runtime compliance, rather than deterministically favoring \textit{ZEBRIS}. A larger $\epsilon_{i,j}(t)$ statistically improves authentication, policy consistency, and SLA satisfaction.}. 

We compare \textit{ZEBRIS} with five baselines\footnote{For fairness, all non-heuristic methods use the same resource-discretized clearing routine when applicable. For methods without deposit-refund regulation, we set $\Delta_{i,j}(t)=0$, so compliance outcomes depend only on seller posture and runtime randomness.}: 
\textit{(i) ResOnly}, a resource-only trading benchmark inspired by \cite{wang2025truthful} that clears trades using raw valuation--ask comparison without privacy risk, compliance cost, or ex-post settlement; 
\textit{(ii) PAware}, a privacy-aware clearing benchmark inspired by \cite{zhuang2025privacy} that considers delay and privacy penalties but removes deposit-refund regulation; 
\textit{(iii) ZTOnly}, a zero-trust-aware benchmark inspired by \cite{zheng2025eo} that considers verification overhead and compliance cost only during ex-ante clearing, but removes ex-post deposit-refund settlement, buyer compensation, and posture feedback;
\textit{(iv) AskFirst}, a cost-oriented heuristic inspired by \cite{wu2025effective} that prioritizes feasible packages with lower effective asks; and 
\textit{(v) ZEBRIS-S}, an ablation variant that preserves zero-trust-aware clearing and deposit-refund settlement but keeps seller posture static over time. 
We evaluate all methods using six metrics: 
\textit{(i) Social welfare (SW)}, the total effective bilateral surplus over accepted trades\footnote{Transfer terms such as payment and deposit redistribution are not counted in SW, but deposit-refund regulation can indirectly affect SW through runtime compliance, seller posture, and future feasibility.}; 
\textit{(ii) Accepted trading ratio (ATR)}, the fraction of active requests admitted into trading; 
\textit{(iii) Average end-to-end delay (AED)}, the mean realized delay of accepted trades; 
\textit{(iv) Average privacy-risk-weighted cost (APRC)}, the average privacy penalty $\beta_i\xi_{i,j}(t)$; 
\textit{(v) Average compliance score (ACS)}, the average score aggregated from authentication success, policy consistency, and SLA satisfaction; and 
\textit{(vi) Seller utility (SU)}, the average realized seller utility after settlement.
	
	\subsection{Performance Evaluation}

\begin{figure}[t]
		\centering
		\setlength{\abovecaptionskip}{-2 mm}
		\includegraphics[width=1\columnwidth]{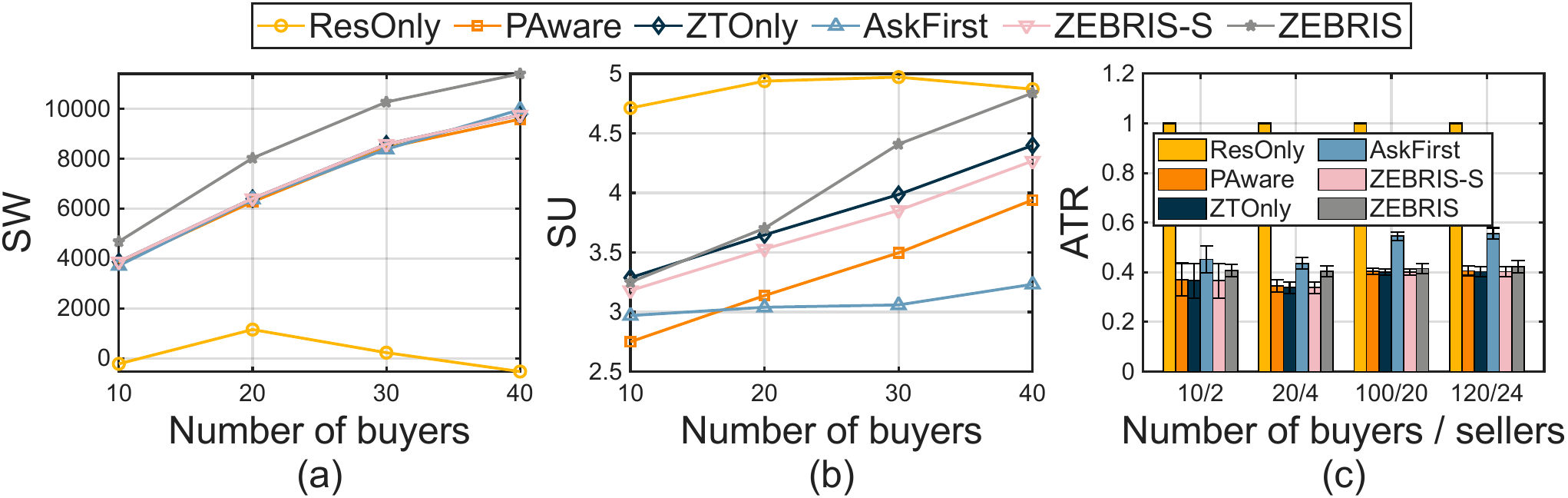}
		\caption{Economic performance and trading behavior under different market scales: (a) SW, (b) SU, and (c) ATR.}
		\label{fig:welfare_utility_acceptance}
		\vspace{-0.4cm}
	\end{figure}

	We first evaluate economic performance and trading behavior in Fig.~\ref{fig:welfare_utility_acceptance}. In Fig.~\ref{fig:welfare_utility_acceptance}(a), the SW of most compared schemes increases with the number of buyers, since a larger buyer population creates more candidate trades and more opportunities for profitable matching. Among all methods, ZEBRIS consistently achieves the highest SW, demonstrating the benefit of jointly integrating zero-trust-compliant package selection, ex-post deposit-refund settlement, and cross-round seller-posture evolution. This also indicates that admitting more trades does not necessarily improve effective welfare, because trades with weak compliance, high privacy exposure, or excessive service delay may reduce the realized market quality. By contrast, ResOnly performs poorly because it admits trades without explicitly accounting for verification overhead, privacy loss, or security mismatch.
	Fig.~\ref{fig:welfare_utility_acceptance}(b) reports SU under different buyer populations. ZEBRIS increases steadily and remains among the best-performing methods. More importantly, it consistently outperforms {ZEBRIS-S}, highlighting the value of feeding realized compliance back to future seller posture. Although {ResOnly} may obtain relatively high SU in some settings, this gain comes from weakly regulated and seller-favorable trading, accompanied by inferior welfare and compliance quality.
	Fig.~\ref{fig:welfare_utility_acceptance}(c) shows that {ResOnly} attains the highest ATR due to aggressive admission, while {ZEBRIS} maintains a moderate ATR by prioritizing effective and trustworthy trades over admitted-request quantity.
	\begin{figure}[t]
		\centering
		\setlength{\abovecaptionskip}{-2 mm}
		\includegraphics[width=1\columnwidth]{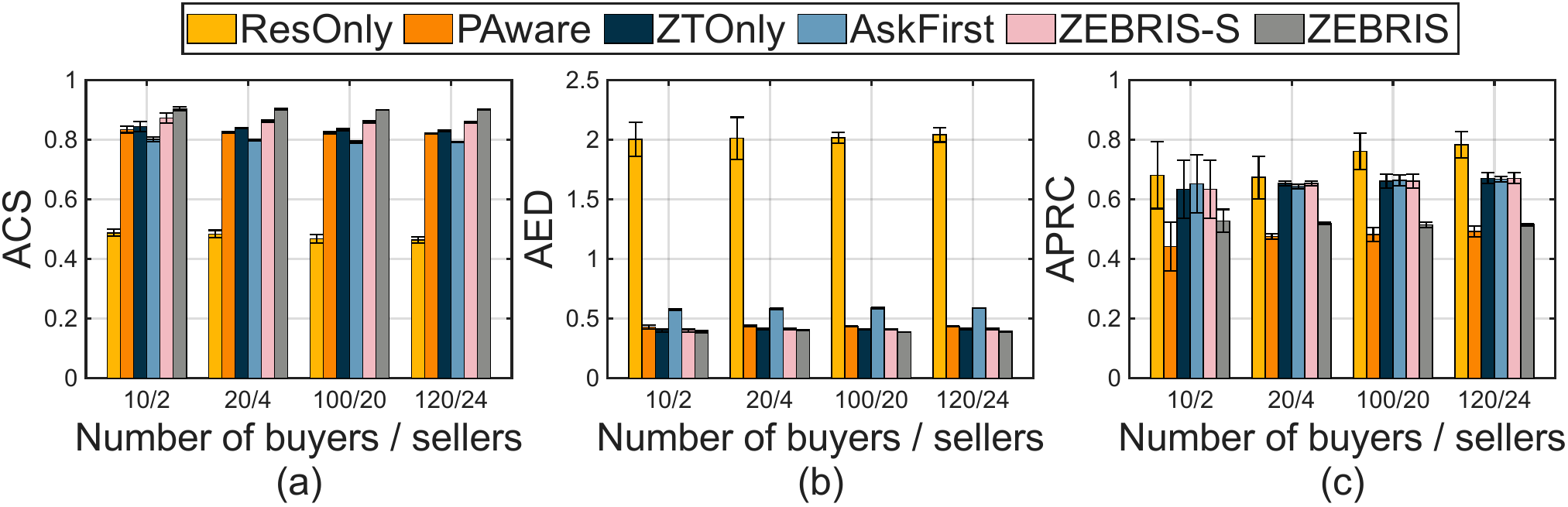}
		\caption{Compliance and service quality under different market scales: (a) ACS, (b) AED, and (c) APRC.}
		\label{fig:compliance_delay_privacy}
		\vspace{-0.5cm}
	\end{figure}
	
	We next evaluate compliance and service quality in Fig.~\ref{fig:compliance_delay_privacy}. In Fig.~\ref{fig:compliance_delay_privacy}(a), ZEBRIS consistently achieves the highest ACS across all market scales. This verifies that coupling ex-post settlement with measurable runtime compliance can effectively discipline seller behavior after clearing. The consistent gain over ZEBRIS-S further indicates that dynamic posture evolution is important for sustaining long-term compliance robustness rather than only improving one-shot execution quality. By contrast, ResOnly obtains the lowest ACS due to the lack of explicit zero-trust regulation and ex-post discipline.
	Fig.~\ref{fig:compliance_delay_privacy}(b) presents the AED. ZEBRIS yields the lowest delay, mainly because package selection explicitly accounts for delay feasibility and because posture feedback gradually favors sellers with more reliable runtime fulfillment. In contrast, aggressive admission without accounting for actual service quality causes ResOnly to suffer the largest delay.
	Fig.~\ref{fig:compliance_delay_privacy}(c) reports the APRC. ZEBRIS consistently achieves the lowest APRC, confirming its advantage in privacy-sensitive edge markets. In contrast, {ZTOnly}, {AskFirst}, and {ZEBRIS-S} incur higher privacy-related cost, while ResOnly performs the worst because privacy risk and zero-trust compliance are ignored during clearing. These results confirm that privacy risk and runtime compliance should be embedded into both ex-ante trading and ex-post settlement.
	
	Overall, the results in Figs.~\ref{fig:welfare_utility_acceptance} and~\ref{fig:compliance_delay_privacy} verify the effectiveness of ZEBRIS from both economic and service-quality perspectives. Compared with representative baselines, ZEBRIS achieves higher SW, stronger ACS, lower AED, and lower APRC, while still maintaining favorable seller utility. More importantly, the comparison with ZEBRIS-S highlights that cross-round seller-posture evolution is not merely an auxiliary refinement, but a key component for stabilizing trustworthy bilateral edge service trading under zero trust.
	
	\section{Conclusion}
	This paper investigated privacy-sensitive edge service trading in dynamic zero-trust edge markets and proposed ZEBRIS, a bilateral edge trading framework for runtime compliance regulation. By modeling edge provisioning as the trading of zero-trust-compliant service packages, ZEBRIS jointly integrates delay penalty, privacy risk, resource cost, and compliance cost into ex-ante clearing, and further disciplines seller-side runtime behavior through measurable ex-post settlement and cross-round security-posture evolution. Experiments showed that ZEBRIS improves social welfare and compliance robustness while reducing service delay and privacy-risk-weighted cost. Future work will extend ZEBRIS to multi-platform edge markets, richer zero-trust policy models, and more adaptive long-term incentive regulation.
	
	\begin{spacing}{0.93}
		\bibliographystyle{ieeetr}
		\bibliography{reference}
	\end{spacing}
	
\end{document}